\documentclass[conference]{IEEEtran}
\ifCLASSINFOpdf
\else
\fi

\usepackage{graphicx}
\usepackage[hyphens]{url}
\usepackage{hyperref}

\usepackage{amsmath}
\usepackage{amssymb}
\usepackage{amsfonts}
\usepackage{amsthm}
\usepackage{textcomp}
\usepackage{bmpsize}
\usepackage[table,dvipsnames]{xcolor}
\usepackage{ifthen}
\usepackage[most]{tcolorbox}
\usepackage{graphicx}
\usepackage[justification=centering]{caption}
\usepackage{multirow,diagbox,tabularx,blindtext}
\usepackage{tabularx}
\usepackage{mdframed}
\usepackage{balance}
\usepackage[utf8]{inputenc}
\usepackage{listings}
\usepackage{array}
\usepackage{makecell}
\usepackage{epigraph}
\usepackage{rotating}
\usepackage{tabularx}
\usepackage[labelfont=bf]{caption} 
\usepackage{threeparttable}
\usepackage{float}
\usepackage{pifont}
\usepackage{stfloats}
\usepackage{longtable}
\usepackage{siunitx}
\usepackage{array}
\usepackage{adjustbox}
\usepackage[htt]{hyphenat}
\usepackage{makecell}
\usepackage{tikz}
\usetikzlibrary{decorations.pathreplacing}

\usepackage[ruled,vlined]{algorithm2e}
\newboolean{showcomments}
\setboolean{showcomments}{true}
\ifthenelse{\boolean{showcomments}}
{ \newcommand{\mynote}[2]{
		\fbox{\bfseries\sffamily\scriptsize#1}
		{\small$\blacktriangleright$\textsf{\emph{#2}}$\blacktriangleleft$}}}
{ \newcommand{\mynote}[2]{}}

\newcounter{findings}
\makeatletter
\newcommand*{\rom}[1]{\expandafter\@slowromancap\romannumeral #1@}
\makeatother
\newcommand{\drebin}[0]{\textsc{DREBIN}\xspace}
\newcommand{\reveal}[0]{\textsc{RevealDroid}\xspace}
\newcommand{\mamaF}[0]{\textsc{MaMaDroid Family}\xspace}
\newcommand{\mamaP}[0]{\textsc{MaMaDroid Package}\xspace}
\newcommand{\mama}[0]{\textsc{MaMaDroid}\xspace}
\newcommand{\malscanA}[0]{\textsc{MalScan Average}\xspace}
\newcommand{\malscanH}[0]{\textsc{MalScan Harmonic}\xspace}
\newcommand{\malscanK}[0]{\textsc{MalScan Katz}\xspace}
\newcommand{\malscanCL}[0]{\textsc{MalScan Closeness}\xspace}
\newcommand{\malscanD}[0]{\textsc{MalScan Degree}\xspace}
\newcommand{\malscanCO}[0]{\textsc{MalScan Concatenate}\xspace}
\newcommand{\malscan}[0]{\textsc{MalScan}\xspace}
\newcommand{\az}[0]{\textsc{AndroZoo}\xspace}

\newcommand{\contrastive}[0]{\textsc{Supervised Contrastive Learning}\xspace}
\newcommand{\guidedRetraining}[0]{\textsc{Guided Retraining}\xspace}

\newcommand{\highlight}[1]{\begin{tcolorbox}[leftrule=1mm,rightrule=1mm,toprule=0mm,bottomrule=0mm,left=2pt,right=2pt,top=1pt,bottom=1pt]
#1
\end{tcolorbox}
}

\newcolumntype{R}[2]{%
    >{\adjustbox{angle=#1,lap=\width-(#2)}\bgroup}%
    l%
    <{\egroup}%
}

\newcolumntype{P}[1]{>{\centering\arraybackslash}p{#1}}
\newcolumntype{M}[1]{>{\centering\arraybackslash}m{#1}}

\begin{document}
%

\title{A two-steps approach to improve the performance of Android malware detectors}

\author{
\IEEEauthorblockN{Nadia Daoudi, Kevin Allix, Tegawendé F. Bissyandé and Jacques Klein}
\IEEEauthorblockA{SnT, University of Luxembourg\\
firstname.lastname@uni.lu}}

\IEEEoverridecommandlockouts

\maketitle

\begin{abstract}
The popularity of Android OS has made it an appealing target to malware developers.
To evade detection, including by ML-based techniques, attackers invest in creating malware that closely resemble legitimate apps.
In this paper, we propose \guidedRetraining, a supervised representation learning-based method that boosts the performance of a malware detector.
First, the dataset is split into ``easy'' and ``difficult'' samples, where difficulty is associated to the prediction probabilities yielded by a malware detector: for difficult samples, the probabilities are such that the classifier is not confident on the predictions, which have high error rates.
Then, we apply our \guidedRetraining method on the difficult samples to improve their classification.
For the subset of ``easy'' samples, the base malware detector is used to make the final predictions since the error rate on that subset is low by construction.
For the subset of ``difficult'' samples, we rely on \guidedRetraining, which leverages the correct predictions and the errors made by the base malware detector to guide the retraining process.
\guidedRetraining focuses on the difficult samples: it learns new embeddings of these samples using Supervised Contrastive Learning and trains an auxiliary classifier for the final predictions.
We validate our method on four state-of-the-art Android malware detection approaches using over 265k malware and benign apps, and we demonstrate that \guidedRetraining can reduce up to 40.41\% prediction errors made by the malware detectors.
Our method is generic and designed to enhance the classification performance on a binary classification task.
Consequently, it can be applied to other classification problems beyond Android malware detection.

\end{abstract}

\section{Introduction} \label{section:introduction}
Android malware plays hide and seek with mobile applications markets operators.
Indeed, new emerging malware apps are increasingly sophisticated~\cite{info12050185, 10.1145/3357384.3357875, 10.1145/3447548.3467168} and challenge state-of-the-art detection techniques, in particular literature ML-based approaches.
These malware apps are designed to closely resemble benign apps in order to hide their malicious behaviour and evade detection.
In typical ML-based malware detection schemes, Android apps are represented using feature vectors (i.e., apps are embedded), which are fed to an algorithm that learns to distinguish malware and benign samples.
In such an embedding space, some malware (or benign) samples occupy a distinct region of the input space~\cite{zhu2018android}.
These samples share similar feature vectors that make them easily distinguishable and separable from the benign (respectively malware) apps in the embedding space.
Nevertheless, there are other malware apps which have feature vectors that are similar to feature vectors of benign samples.
Such apps are located in regions of the embedding space where malware and benign samples are not perfectly separable and distinguishable.
In such regions, malware and benign apps overlap, which leads to misclassifications.

Deep representation learning aims to extract relevant patterns from the input data and discard the noise.
Several techniques~\cite{supContrastive, 7933051, 8987459, huang2019supervised, walmsley2021practical} have leveraged the class labels to generate powerful representations, which has led to state-of-the-art performance.
Indeed, supervised representation learning methods are trained to automatically learn characteristic features of samples that share the same class labels.
The resulting embeddings would be passed to a classifier that can map the samples to their respective classes.
Recently, \contrastive~\cite{supContrastive} has been proposed to maximise the embedding similarity of samples from the same class and minimise the embedding similarity of samples belonging to different classes.
This representation learning method transforms the input data into an embedding space in which samples with the same labels are close to each other, so they can have similar representations.
Furthermore, it increases the distance between samples from different classes so they can get distinct representations.
\contrastive seems to propose a solution for overlapping malware and benign samples since it transforms the input data into a new embedding space in which samples from the same class are grouped together and separated from the other class.

In binary classification, we can distinguish between two categories of samples based on their input labels: positives and negatives (i.e., malware and benign).
It is also possible to classify samples into \emph{easy} and \emph{difficult} instances based on their feature vectors.
Easy samples refer to positive and negative instances which a classifier can easily identify and correctly predict their classes.
The difficult samples can also be positives or negatives, but they have similar input features that make it challenging for the classifier to correctly identify their classes.
For a base classifier, identifying the class of the easy samples would be straightforward, which results in low prediction errors.
As for the difficult samples, they would need more advanced techniques to better discriminate the two classes.

In this paper, we propose to address the problem of malware detection in two steps:
The first step of the classification would contain the samples that are easy to predict by a base classifier.
We rely on the prediction probabilities of the base classifier to decide whether a sample is easy or difficult.
Moreover, all the samples that are identified to be easy would be predicted by that classifier.
If a sample is tagged as difficult, it would not be predicted by the base classifier but passed to the second step in which it will be handed over to an auxiliary classifier trained via our \guidedRetraining method.
As its name suggests, our technique is designed to guide the retraining on the difficult samples to reduce the prediction errors.
We rely on the predictions generated by the base classifier on the difficult samples to learn distinctive representations for each class.
Specifically, we leverage supervised \contrastive learning to generate embeddings for the difficult samples in five guided steps that teach the model to learn from the correct predictions and errors made by the base classifier.
Then, we train an auxiliary classifier on the generated embeddings so it can make the final classification decision on the difficult samples.
In short, to predict the class of a given sample, we check whether it is easy or difficult to predict by the base classifier.
If it is easy, the prediction decision of the base classifier is taken into account. 
Otherwise, it will be classified by the auxiliary classifier.

To validate the effectiveness of our method, we evaluate it on four state-of-the-art Android malware detectors (i.e., with their variants) that have been successfully replicated in the literature~\cite{reproduction}: \drebin~\cite{drebin}, \reveal~\cite{revealdroid}, \mama~\cite{mamadroid:ndss}, and \malscan~\cite{malscan}.
These detectors consider various features to discriminate between malware and benign apps, and they have been reported to be highly effective.
Our experiments demonstrate that the prediction errors made by state-of-the-art Android malware detectors can be reduced via our \guidedRetraining method.
Specifically, we show that our technique boosts the detection performance for nine out of ten malware detectors and reduces up to 40.41\% prediction errors made by the classifiers.

Overall, our work makes the following contributions:

\begin{itemize}
    \item We propose to address the malware detection problem in two steps: the first step deals with the detection of the easy samples, and the second step is intended for the difficult apps
    \item We design a new technique, \guidedRetraining, that improves the classification of the difficult apps
    \item We validate the effectiveness of our method on four state-of-the-art Android malware detectors
\end{itemize}




\section{Approach}\label{sec:approach}
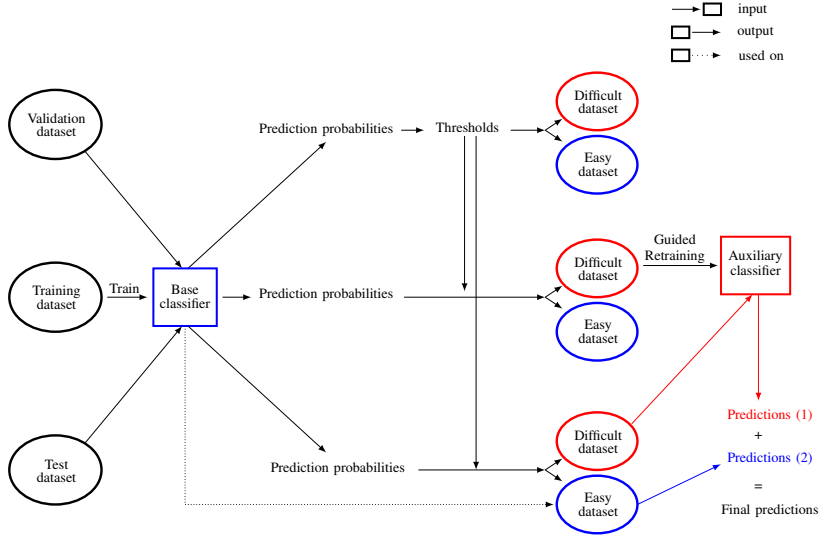
\begin{figure*}[!htbp]
    \centering
    \resizebox{0.6\linewidth}{!}{
    \begin{tikzpicture}
\tikzstyle{arrowStyle}=[-latex]
\tikzset{node/.style={rounded corners, align=center,scale=1}}

    \begin{scope}[cm={1,0,0,1,(0, 0)}]
    
    \draw[black, very thick] (0,0.4) ellipse (.8 and 0.6);
    \draw (0.02, 0.5) node[]{\scriptsize Validation};
    \draw (0.02, 0.28) node[]{\scriptsize dataset};

    \draw[black, ->,>=latex] (0.5, -.05) -- (2.2, -2.1);
    \draw[black, ->,>=latex] (2.3, -2.1) -- (4.7, .1);

    \draw (4.7, 0.31) node[]{\scriptsize Prediction probabilities};
    
    \draw[->,>=latex] (6., .3) -- (6.4, .3);
    
    \draw (7.15, 0.35) node[]{\scriptsize Thresholds};
    \draw[->,>=latex] (7.9, .3) -- (8.5, .3);
    
    \draw[->,>=latex] (7.1, .2) -- (7.1, -2.5);
    \draw[->,>=latex] (7.3, .2) -- (7.3, -5.6);

    \draw[->,>=latex] (8.5, .3) -- (8.8, .5);
    \draw[->,>=latex] (8.5, .3) -- (8.8, .1);
    
    \draw[red, very thick] (9.4, .8) ellipse (.7 and 0.5);
    \draw (9.42, 0.9) node[]{\scriptsize Difficult};
    \draw (9.42, 0.68) node[]{\scriptsize dataset};

    \draw[blue, very thick] (9.4, -.3) ellipse (.7 and 0.5);
    \draw (9.42, -.2) node[]{\scriptsize Easy};
    \draw (9.42, -.42) node[]{\scriptsize dataset};

    \draw[->,>=latex] (10.7, 2.4) -- (11.2, 2.4);
    \draw [black, line width=1] (11.25, 2.3) rectangle (11.55, 2.5);
    \draw (12.1, 2.4) node[]{\scriptsize input};
    
    \draw [black, line width=1] (10.7, 2.1) rectangle (11., 1.9);
    \draw[->,>=latex] (11.05, 2.) -- (11.55, 2.);
    \draw (12.15, 2.) node[]{\scriptsize output};
    
    \draw [black, line width=1] (10.7, 1.7) rectangle (11., 1.5);
    \draw[dotted, ->,>=latex] (11.05, 1.6) -- (11.55, 1.6);
    \draw (12.25, 1.6) node[]{\scriptsize used on};
    
    \end{scope}
    \begin{scope}[cm={1,0,0,1,(0, -3)}]
    
    \draw[black, very thick] (0,0.4) ellipse (.8 and 0.6);
    \draw (0.02, 0.5) node[]{\scriptsize Training};
    \draw (0.02, 0.28) node[]{\scriptsize dataset};
    
    \draw[->,>=latex] (.9, .4) -- (1.6, .4);
    \draw (1.2, 0.55) node[]{\scriptsize Train};
    
    \draw [blue, line width=1] (1.7, -0.1) rectangle (2.8, .9); 
    \draw (2.25, 0.53) node[]{\scriptsize Base};
    \draw (2.25, 0.29) node[]{\scriptsize classifier};

    \draw[black, ->,>=latex] (2.9, .4) -- (3.4, .4);
    \draw (4.7, 0.44) node[]{\scriptsize Prediction probabilities};
    

    \draw[->,>=latex] (6.05, .4) -- (8.5, .4);
    
    \draw[->,>=latex] (8.5, .4) -- (8.8, .6);
    \draw[->,>=latex] (8.5, .4) -- (8.8, .2);

    \draw[red, very thick] (9.4, .9) ellipse (.7 and 0.5);
    \draw (9.42, 1.) node[]{\scriptsize Difficult};
    \draw (9.42, 0.78) node[]{\scriptsize dataset};
    
    \draw[blue, very thick] (9.4, -.2) ellipse (.7 and 0.5);
    \draw (9.42, -.1) node[]{\scriptsize Easy};
    \draw (9.42, -.32) node[]{\scriptsize dataset};

    \draw[->,>=latex] (10.2, .95) -- (11.5, .95);
    \draw (10.75, 1.4) node[]{\scriptsize Guided};
    \draw (10.75, 1.1) node[]{\scriptsize Retraining};
    
    \draw [red, line width=1] (11.55, 1.45) rectangle (12.75, .45);
    \draw (12.15, 1.1) node[]{\scriptsize Auxiliary};
    \draw (12.15, 0.86) node[]{\scriptsize classifier};

    \end{scope}
    \begin{scope}[cm={1,0,0,1,(0, -6)}]
    
    \draw[black, very thick] (0,0.4) ellipse (.8 and 0.6);
    \draw (0.02, 0.5) node[]{\scriptsize Test};
    \draw (0.02, 0.28) node[]{\scriptsize dataset};
    
    \draw[black, ->,>=latex] (0.5, .85) -- (2.2, 2.9);
    \draw[black, ->,>=latex] (2.3, 2.9) -- (4.7, .7);
    
    \draw (4.9, 0.43) node[]{\scriptsize Prediction probabilities};
    
    \draw[->,>=latex] (6.3, .4) -- (8.5, .4);
    \draw[->,>=latex] (8.5, .4) -- (8.8, .6);
    \draw[->,>=latex] (8.5, .4) -- (8.8, .2);
    
     \draw[red, very thick] (9.4, .9) ellipse (.7 and 0.5);
     \draw (9.42, 1.) node[]{\scriptsize Difficult};
     \draw (9.42, 0.78) node[]{\scriptsize dataset};
    
     \draw[blue, very thick] (9.4, -.2) ellipse (.7 and 0.5);
     \draw (9.42, -.1) node[]{\scriptsize Easy};
     \draw (9.42, -.32) node[]{\scriptsize dataset};
    
     \draw[red, ->,>=latex] (10., 1.2) -- (12.1, 3.45);
     \draw[red, ->,>=latex] (12.2, 3.45) -- (12.2, 1.6);
     
    \draw[densely dotted, ,>=latex] (2.25, 2.9) -- (2.25, -.2);
    \draw[densely dotted, ->,>=latex] (2.25, -.2) -- (8.65, -.2);
    
    \draw[blue, ->,>=latex] (10.1, -.2) -- (11.55, .5);

    \draw (12.4, 1.35) node[red]{\scriptsize Predictions (1)};
    \draw (12.2, 1.) node[]{\scriptsize +};
    \draw (12.4, .6) node[blue]{\scriptsize Predictions (2)};
    \draw (12.2, .1) node[]{\scriptsize =};
    \draw (12.4, -.3) node[]{\scriptsize Final predictions};
    
    \end{scope}
    
\end{tikzpicture}}
    \caption{An overview of our approach}
    \label{fig:overview}
\end{figure*}

Our method aims to leverage deep learning techniques in order to boost the performance of a binary base classifier.
We present in Fig.~\ref{fig:overview} an overview of our method.
The first step consists of training a base classifier on the whole dataset. 
Then, we leverage the prediction probabilities of the base classifier to split the dataset into two subsets: easy and difficult samples.
The difficult samples are used to train an auxiliary classifier via our \guidedRetraining method. 
Given a new sample, if it is identified as an easy sample, it will be predicted by the base classifier. 
Otherwise, the prediction decision will be made by the auxiliary classifier that is trained on the difficult samples via our \guidedRetraining method.
In the following, we describe the main steps of our approach which are: The base classifier training, Difficult samples identification, and \guidedRetraining.

\subsection{The base classifier training}\label{sec:approach:step1}
Our approach is designed to boost the performance of an existing binary classifier that we denote as the base classifier.
The type of this classifier is not important, but ideally it should be able to output the prediction probabilities, i.e., not only a binary classification (such as \emph{malware} or \emph{benign}) but a value, typically between 0 and 1, that indicates the likelihood that a given sample is a malware.
If the classifier does not generate prediction probabilities, we propose other solutions in Section~\ref{sec:evaluation_setup:setup}.

The first step consists of splitting up the dataset into three subsets: training, validation, and test.
We train the base classifier using all the samples in the training subset.
Then, the base classifier is used to assign a probability of prediction and a binary prediction to each sample in the dataset (i.e., samples that belong to the training, validation and test datasets).

\subsection{Difficult samples identification}\label{sec:approach:step2}
The aim of this step is to identify the samples that are ``difficult'' to predict by the base classifier.
The criteria we use to identify these samples is their probabilities of prediction.

In a binary classification experiment, if the model is confident about the label of a given sample, it assigns a high probability of prediction to the class that is associated with that label (i.e., a probability of prediction that is close to 1).
Otherwise, the two classes get similar probabilities of prediction (i.e., the probabilities of prediction for the two classes are close to 0.5).
The predicted labels are then decided based on the probabilities of predictions.
Generally, when the probability of prediction for the positive class (or the negative class) is higher than 0.5, the classifier predicts the sample as positive (or negative).
Since the probability of prediction for the negative class can be deduced from the probability of prediction of the positive class (i.e., the two probabilities sum up to 1), we consider only the probability of prediction of the positive class in the following, and we denote it $p$.

In our approach, we leverage the probabilities of prediction to split a dataset into easy and difficult subsets.
After it is trained, the base classifier would assign either a very high or a very small probability of prediction $p$ to the samples that it can predict their labels with a high confidence.
Specifically, if $p$ is very high, the base classifier is confident that the sample belongs to the positive class.
Conversely, if $p$ is very low, the classifier is confident that the sample belongs to the negative class.
If a given sample is attributed a very high or a very small probability of prediction, we consider that it is an easy sample.
Otherwise, it is considered to belong to the difficult subset.


\subsubsection{Identifying the probability thresholds}
From the previous step, our base classifier has attributed a probability of prediction to each sample in the whole dataset.
The next step consists of tagging each sample in the dataset as easy or difficult based on its probability of prediction.
To this end, we need to identify two probability of prediction thresholds for considering a sample as easy or difficult.
Specifically, we rely on the first probability threshold to decide whether the prediction probability $p$ of a given sample is high enough to consider that sample as easy (i.e., in this case the sample is an easy positive since $p$ is high).
Similarly, when the prediction probability $p$ of a given sample is small, we need another probability threshold to decide whether $p$ is small enough to tag the sample as easy (i.e., in this case the sample is an easy negative).

We rely on the validation dataset to determine the values of the two probability of prediction thresholds.
Specifically, since the validation samples are classified into TNs (i.e., True Negatives), FPs (i.e., False Positives), FNs (i.e., False Negatives), and TPs (i.e., True Positives), we determine the probability thresholds that satisfy the following constraints:
\begin{itemize}
    \item The probability threshold for considering a sample as an easy positive must ensure that the number of the false positives in the easy validation dataset is equal to $X\%$ of the total number of FPs (i.e., FPs predicted by the base classifier on the whole validation dataset). We denote this threshold $th_p$.
    \item The probability threshold for classifying a sample as an easy negative must guarantee that the number of the false negatives in the easy validation dataset is equal to $Y\%$ of the total number of FNs (i.e., FNs predicted by the base classifier based on the whole validation dataset). We denote this threshold $th_n$.
\end{itemize}

To identify the values of the two probability thresholds, we need to compute the number of FPs and FNs that we tolerate in the easy validation dataset. 
We note these variables $toleratedFPs$ and $toleratedFNs$ and we calculate their values as follows:

\[ toleratedFPs = \frac{X \times FP_v}{100} \]

\[ toleratedFNs = \frac{Y \times FN_v}{100} \]

where $FP_v$ and $FN_v$ represent the number of FPs and FNs returned by the base classifier on the whole validation dataset respectively.

The process of identifying the two probability of prediction thresholds is adequately detailed in Algorithm~\ref{algo:thresholds_selection}.

\begin{algorithm}[!ht]
\caption{Thresholds selection}
\textbf{Input:} vDataset, yProbabilities, toleratedFPs, toleratedFNs, indicesOfFPs, indicesOfFNs \\
\textbf{Output:} thresholdFPs, thresholdFNs \\
\smallskip

counterFPs $\gets 0$ \\
counterFNs $\gets 0$ \\
lenData $\gets vDataset.length() $ \\

probasIndicesPos $\gets \emptyset$ \\
probasIndicesNeg $\gets \emptyset$ \\

\For{$i \leftarrow 1, lenData$}{
    \uIf{yProbabilities(i) $\geq$ 0.5}{
     probasIndicesPos $\gets$ probasIndicesPos + (yProbabilities(i), i) } 
     // We keep track of the index of the sample to verify that it is not among the FPs and FNs \\
     // We later search the index in indicesOfFPs and indicesOfFNs lists \\
    \Else{probasIndicesNeg $\gets$ probasIndicesNeg + (yProbabilities(i), i)}}
    
probasIndicesPos $\gets$ probasIndicesPos.inverselySortProbas() \\
// The prediction probabilities of the positive samples are sorted in descending order \\
probasIndicesNeg $\gets$ probasIndicesNeg.sortProbas() \\
// The prediction probabilities of the negative samples are sorted in ascending order \\

lenPos $\gets probasIndicesPos.length() $ \\
lenNeg $\gets probasIndicesNeg.length() $ \\

\For{$i \leftarrow 1, lenPos$}{
    \uIf{counterFPs == toleratedFPs}{
    thresholdFPs $\gets$ probasIndicesPos[i][0] \\
    break}
    
    \uIf{probasIndicesPos[i][1] in indicesOfFPs}{
    counterFPs $\gets$ counterFPs + 1}}

\For{$i \leftarrow 1, lenNeg$}{
    \uIf{counterFNs == toleratedFNs}{
    thresholdFNs $\gets$ probasIndicesNeg[i][0] \\
    break}
    
    \uIf{probasIndicesNeg[i][1] in indicesOfFNs}{
    counterFNs $\gets$ counterFNs + 1}}

\label{algo:thresholds_selection}
\end{algorithm}

The inputs to this algorithm are the validation dataset, the probabilities of prediction returned by the base classifier on the validation dataset, toleratedFPs, toleratedFNs, and the indices of the $FP_v$ and $FN_v$ in the validation dataset (i.e., we consider that each instance in the dataset has a unique index, and we denote the lists of the FPs and FNs indices as indicesOfFPs and indicesOfFNs respectively).
To identify the threshold of the positives, we first select all the samples from the validation dataset that have their $p$ $\geq$ 0.5 and we sort their probabilities in descending order.
We also keep track of the indices of these samples in the validation dataset to verify whether they are predicted as TPs or FPs by the base classifier (i.e., based on indicesOfFPs list).
Then, we initialise a counter of the number of FPs in the easy dataset and we iterate over the sorted samples starting from the one with the highest probability of prediction.
During each iteration, we first check whether the value of the FPs counter has reached the number of toleratedFPs, in which case we stop the iteration and set the $th_p$ to the current probability of prediction.
Otherwise, we increment the counter of FPs if the sample has been predicted as FP by the base classifier.

We apply the same technique to identify the value of negatives threshold $th_n$.
We select the samples that have their $p$ $\leq$ 0.5 and we sort their probabilities in ascending order since the classifier is confident about the samples with low probabilities of prediction.
Similarly, we keep a counter for the number of FNs that are tolerated in the easy dataset and we iterate over the sorted samples starting from the one with the lowest probability of prediction.
When the value of the FNs counter is equal to the value of toleratedFNs, we stop the iteration.
We then set the value of $th_n$ to the probability of prediction of the last sample in which the iteration stopped.

\subsubsection{Splitting the datasets}
After identifying the values of $th_p$ and $th_n$, we split our datasets into easy and difficult subsets.



 

The easy dataset contains all the samples whose probabilities of prediction satisfy:


\begin{multline*}
    easyDataset = \{ x_i \in dataset \mid 
     th_n \geq p_\mathrm{i}  \mbox{~or~}  th_p \leq p_\mathrm{i} \}
\end{multline*}
where $p_\mathrm{i}$ represents the probability of prediction for the sample $x_i$.

The easy dataset includes all the positive samples whose prediction probabilities are greater than the threshold $th_p$ (i.e., they are predicted as positives with high confidence by the base classifier).
It also includes the negative samples whose prediction probabilities are smaller than the threshold $th_n$ (i.e., they are predicted as negatives with high confidence by the base classifier).

As for the difficult dataset, it contains all the samples that do not satisfy the constraints of the easy dataset.
Specifically, it includes the samples whose prediction probabilities are at the same time below the threshold $th_p$ and above the threshold $th_n$ (i.e., the base classifier is not confident that these samples are positives or negatives).
The samples in the difficult dataset satisfy:
\begin{multline*}
    difficultDataset = \{ x_i \in dataset \mid 
      th_n < p_\mathrm{i} < th_p \}
\end{multline*}

At the end of this step, we have the training, validation, and test datasets split into easy and difficult subsets.

\subsection{\guidedRetraining}\label{sec:approach:step3}
In our approach we make use of \contrastive~\cite{supContrastive} to generate the embeddings of the difficult samples.
This method aims to represent the dataset in such a way that samples belonging to the same class are close to each other in the embedding space. Similarly, the samples belonging to different classes are far from each other in the embedding space.
\contrastive works in two stages: First, it generates the embeddings using an Encoder followed by a Projection Network (we refer to both of them as the \texttt{Model}).
After the training is done, the Projection Network is discarded and a classifier is trained on the embeddings from the last layer of the Encoder. 
This classifier is referred to as the auxiliary classifier.
At the end of the second stage, the samples are classified into their respective classes.
Using \contrastive, we aim to create contrasted representations for the samples in the difficult subsets which would help to better classify them into their respective classes.

From the previous step (i.e., Section~\ref{sec:approach:step2}), we have created two validation subsets: easy and difficult.
By construction, the difficult validation subset contains most of the misclassifications made by the base classifier.
Specifically, it contains (100 - X)\% of the total number of FPs contained on the whole validation dataset. 
Likewise, the number of FNs reaches (100 - Y)\% of the total number of FNs in the validation dataset.
The training difficult subset is also expected to include similar proportions of FPs and FNs (i.e., it includes most of the prediction errors from the whole training dataset).
We remind that the difficult subsets also contain correct predictions made by the base classifier.
In the following, we use $TN's_\mathrm{tr}$, $FP's_\mathrm{tr}$, $FN's_\mathrm{tr}$, and $TP's_\mathrm{tr}$ to refer to TNs, FPs, FNs, and TPs of the base classifier on the difficult training subset.


As the title suggests, we propose a method that would guide the retraining on the difficult samples.
Specifically, we aim to help the \texttt{Model} to distinguish between four categories of samples in the difficult training dataset.
These categories are: $TN'_\mathrm{tr}$, $FP'_\mathrm{tr}$, $FN'_\mathrm{tr}$, and $TP'_\mathrm{tr}$.
We present in Figure~\ref{fig:guided_retraining} an overview of our \guidedRetraining approach.

\begin{figure*}[!htbp]
    \centering
    \resizebox{0.7\linewidth}{!}{
    \begin{tikzpicture}
\tikzstyle{arrowStyle}=[-latex]
\tikzset{node/.style={rounded corners, align=center,scale=1}}

    \begin{scope}[cm={1,0,0,1,(1.55, 0)}]
    
    \draw[black, very thick] (0,0.4) ellipse (1.2 and 0.6);
    \draw (0.02, 0.5) node[]{\scriptsize Difficult training};
    \draw (0.02, 0.28) node[]{\scriptsize dataset};
    
    \draw[->,>=latex] (., -.3) -- (., -2.);
    \draw (-.64, -1.1) node[]{\scriptsize Predict};

    \draw [blue, line width=1] (.15, -1.5) rectangle (1.25, -.65); 
    \draw (.7, -.9) node[]{\scriptsize Base};
    \draw (.7, -1.14) node[]{\scriptsize classifier};

    \draw [thick, ->] (-1.25, 0.4) .. controls (-8.5,-1.) and (-8.5,-4.5)  .. (-4.95,-6.4);
    \draw [thick, ->] (-7.2, -4.05) .. controls (-6.4,-6.5) and (-5.4,-7.1)  .. (-1.65,-6.7);
    \draw [thick, ->] (-7.2, -4.05) .. controls (-6.4,-6.9) and (-5.4,-7.3)  .. (1.75,-6.8);
    \draw [thick, ->] (-7.2, -4.05) .. controls (-6.4,-7.1) and (-5.4,-7.5)  .. (5.05,-6.9);
    
    \end{scope}
    
    
    \begin{scope}[cm={1,0,0,1,(0, 0)}]

    \draw[decorate, decoration={brace,amplitude=3pt, raise=4pt},line width=1, yshift=0pt] (-4.9,-2.5) -- (8,-2.5);
    \draw[-,>=latex,line width=1] (-4.8, -2.4) -- (-4.8, -2.6);
    \draw[-,>=latex,line width=1] (-.9, -2.4) -- (-.9, -2.6);
    \draw[-,>=latex,line width=1] (3.8, -2.4) -- (3.8, -2.6);
    \draw[-,>=latex,line width=1] (7.9, -2.4) -- (7.9, -2.6);

    \draw[black, very thick] (-4.7, -3.2) ellipse (.7 and 0.5);
    \draw (-4.68, -3.2) node[]{\scriptsize \bf{TPs}};
    
    \draw[black, very thick] (-.8, -3.2) ellipse (.7 and 0.5);
    \draw (-.78, -3.2) node[]{\scriptsize \bf{FPs}};
    
    \draw[black, very thick] (3.9, -3.2) ellipse (.7 and 0.5);
    \draw (3.92, -3.2) node[]{\scriptsize \bf{TNs}};
    
    \draw[black, very thick] (8., -3.2) ellipse (.7 and 0.5);
    \draw (8.02, -3.2) node[]{\scriptsize \bf{FNs}};

    \draw[-,>=latex,line width=1] (-4.7, -3.75) -- (-3.3, -4.6);
    \draw[-,>=latex,line width=1] (-.8, -3.75) -- (-3.3, -4.6);
    
    \draw[-,>=latex,line width=1] (-4.7, -3.75) -- (., -4.6);
    \draw[-,>=latex,line width=1] (3.9, -3.75) -- (., -4.6);
    
    \draw[-,>=latex,line width=1] (-.8, -3.75) -- (3.4, -4.6);
    \draw[-,>=latex,line width=1] (8., -3.75) -- (3.4, -4.6);
    
    \draw[-,>=latex,line width=1] (3.9, -3.75) -- (6.7, -4.6);
    \draw[-,>=latex,line width=1] (8., -3.75) -- (6.7, -4.6);
    
    \draw[->,>=latex,line width=1] (-3.3, -4.6) -- (-3.3, -5.2);
    \draw (-2.8, -4.8) node[]{\scriptsize Train};
    
    \draw[->,>=latex,line width=1] (-., -4.6) -- (-., -5.2);
    \draw (.5, -4.8) node[]{\scriptsize Train};
    
    \draw[->,>=latex,line width=1] (3.4, -4.6) -- (3.4, -5.2);
    \draw (3.9, -4.8) node[]{\scriptsize Train};
    
    \draw[->,>=latex,line width=1] (6.7, -4.6) -- (6.7, -5.2);
    \draw (7.2, -4.8) node[]{\scriptsize Train};

    \draw [green, line width=1] (-3.8, -6.) rectangle (-2.7, -5.2); 
    \draw (-3.25, -5.6) node[]{\scriptsize \texorpdfstring{Model\textsubscript{1}}{}};
    
    \draw [green, line width=1] (-.5, -6.) rectangle (.6, -5.2); 
    \draw (.05, -5.6) node[]{\scriptsize \texorpdfstring{Model\textsubscript{2}}{}};
    
    \draw [green, line width=1] (2.9, -6.) rectangle (4., -5.2); 
    \draw (3.45, -5.6) node[]{\scriptsize \texorpdfstring{Model\textsubscript{3}}{}};
    
    \draw [green, line width=1] (6.1, -6.) rectangle (7.2, -5.2); 
    \draw (6.65, -5.6) node[]{\scriptsize \texorpdfstring{Model\textsubscript{4}}{}};

    \draw[->,>=latex,line width=1] (-3.3, -6.) -- (-3.3, -7.4);
    \draw (-2.5, -6.2) node[]{\scriptsize Generate};
    \draw (-2.4, -6.45) node[]{\scriptsize embedding};
    
    \draw[->,>=latex,line width=1] (-., -6.) -- (-., -7.4);
    \draw (.85, -6.2) node[]{\scriptsize Generate};
    \draw (.84, -6.45) node[]{\scriptsize embedding};
    
    \draw[->,>=latex,line width=1] (3.4, -6.) -- (3.4, -7.4);
    \draw (4.25, -6.2) node[]{\scriptsize Generate};
    \draw (4.24, -6.45) node[]{\scriptsize embedding};
    
    \draw[->,>=latex,line width=1] (6.7, -6.) -- (6.7, -7.4);
    \draw (7.55, -6.2) node[]{\scriptsize Generate};
    \draw (7.54, -6.45) node[]{\scriptsize embedding};
    
    \draw (-3.2, -7.6) node[]{\scriptsize Embedding 1};
    \draw (0.1, -7.6) node[]{\scriptsize Embedding 2};
    \draw (3.4, -7.6) node[]{\scriptsize Embedding 3};
    \draw (6.8, -7.6) node[]{\scriptsize Embedding 4};
    
    \draw[-,>=latex,line width=1] (-3.2, -7.8) -- (1.9, -8.4);
    \draw[-,>=latex,line width=1] (., -7.8) -- (1.9, -8.4);
    \draw[-,>=latex,line width=1] (3.4, -7.8) -- (1.9, -8.4);
    \draw[-,>=latex,line width=1] (6.8, -7.8) -- (1.9, -8.4);
    
    \draw[->,>=latex,line width=1] (1.9, -8.4) -- (1.9, -9.2);
    \draw (2.9, -8.8) node[]{\scriptsize Concatenation};
    
    \draw (1.9, -9.4) node[]{\scriptsize Embedding};
    
    \draw[->,>=latex,line width=1] (1.9, -9.6) -- (1.9, -10.3);
    \draw (2.4, -9.9) node[]{\scriptsize Train};
    
    \draw [green, line width=1] (1.4, -10.3) rectangle (2.4, -11.1);
    \draw (1.9, -10.7) node[]{\scriptsize \texorpdfstring{Model\textsubscript{5}}{} };
    
    \draw[->,>=latex,line width=1] (1.9, -11.1) -- (1.9, -11.9);
    \draw (2.7, -11.3) node[]{\scriptsize Generate};
    \draw (2.7, -11.55) node[]{\scriptsize embedding};
    
    \draw (1.9, -12.1) node[]{\scriptsize Final embedding};
    
    \draw[->,>=latex,line width=1] (1.9, -12.3) -- (1.9, -13.);
    \draw (2.4, -12.6) node[]{\scriptsize Train};
    
    \draw [red, line width=1] (1.3, -13.) rectangle (2.6, -13.9); 
    \draw (1.9, -13.3) node[]{\scriptsize Auxiliary };
    \draw (1.9, -13.6) node[]{\scriptsize classifier };
    
    \end{scope}
\end{tikzpicture}}
    \caption{An illustration of our \guidedRetraining method}
    \label{fig:guided_retraining}
\end{figure*}

Since training a binary classifier requires a dataset that contains samples from two classes (i.e., positives and negatives), we make use of the different combinations of subsets in the training difficult dataset to help the \texttt{Model} generate more contrasted embeddings.
Specifically, we first train a \texttt{Model} using $TP's_\mathrm{tr}$ (i.e., they have positive real labels) and $FP's_\mathrm{tr}$ (i.e., they have negative real labels), and we denote it \texorpdfstring{\texttt{Model}\textsubscript{1}}{}.
Basically, we guide \texorpdfstring{\texttt{Model}\textsubscript{1}}{} to distinguish between the positive samples that are correctly predicted by the base classifier and the negative samples that are all misclassified by the same classifier.
Consequently, \texorpdfstring{\texttt{Model}\textsubscript{1}}{} focuses on learning a contrasted representation for the true positives and the false positives in the difficult training dataset.
Then, we train another \texttt{Model} using $TN's_\mathrm{tr}$ (i.e., they have negative real labels) and $FN's_\mathrm{tr}$ (i.e., they have positive real labels) and we denoted it \texorpdfstring{\texttt{Model}\textsubscript{2}}{}.
This \texttt{Model} would learn to distinguish between the true negatives and the false negatives predicted by the base classifier on the difficult training subset.
Similarly, we train \texorpdfstring{\texttt{Model}\textsubscript{3}}{} on $TP'_\mathrm{tr}$ (i.e., they have positive real labels) and $TN'_\mathrm{tr}$ (i.e., they have negative real labels), and \texorpdfstring{\texttt{Model}\textsubscript{4}}{} on $FP'_\mathrm{tr}$ (i.e., they have negative real labels) and $FN'_\mathrm{tr}$ (i.e., they have positive real labels).

In summary, the four \texttt{Models} are trained on two difficult training subsets that the base classifier has: (1) either correctly or incorrectly classified both of them, (2) correctly predicted one subset and misclassified the other subset.

After the four \texttt{Models} are trained, they are used to generate embeddings for the difficult training dataset.
Specifically, four embeddings are generated for each sample in the difficult training dataset.
Then, we concatenate the four feature representations of each sample into one vector in order to have one embedding per sample.

To create more contrasted representations for the difficult samples, we train another \texttt{Model} on the concatenated embeddings and we denote it \texorpdfstring{\texttt{Model}\textsubscript{5}}{}.
Basically, \texorpdfstring{\texttt{Model}\textsubscript{5}}{} is trained on all the samples from the difficult training dataset, which would create fine-grained contrasted representations based on the embeddings generated by the four previous \texttt{Models}. 
Indeed, \texorpdfstring{\texttt{Model}\textsubscript{5}}{} would learn from the concatenated embeddings of each sample in the difficult subset (i.e., whether the base classifier has correctly or incorrectly predicted it) to generate the final feature representations.

The last step in our approach is to train the auxiliary classifier on the difficult training embeddings that are generated by \texorpdfstring{\texttt{Model}\textsubscript{5}}{}.
This classifier is trained on all the difficult samples in the training subset.
The final classification decision of the difficult samples is given by the auxiliary classifier.
We remind that for the easy datasets, it is the base classifier that is in charge of predicting their class labels, as illustrated in Fig~\ref{fig:overview}.

\section{Evaluation setup}\label{sec:evaluation_setup}
In this section, we first present the research questions we investigate in our study and the evaluation subjects we use to assess the effectiveness of our approach.
Then, we describe the dataset, the architecture of both the \texttt{Model} and the auxiliary classifier, and we overview our experimental setup.

\subsection{Research questions}\label{sec:evaluation_setup:rqs}
In our study, we investigate the possibility of selecting and separating the samples that are most challenging to classify.
Specifically, we aim to identify the difficult subset in a dataset that would contain most of the prediction errors.

\begin{itemize}
    \item \textbf{RQ1:} To what extent it is feasible to split a dataset into two subsets, one with fewer prediction errors and one with most errors?

    
\end{itemize}


After identifying the difficult subset in a dataset, we investigate the impact of the classic retraining method on the detection performance.
Specifically, we train only one \texttt{Model} on the difficult training dataset to generate the embeddings. Then, we train an auxiliary classifier for the prediction decision.
\begin{itemize}
    \item \textbf{RQ2:} How effective is the classic retraining method in improving the classification on the difficult subset?
\end{itemize}


Finally, we evaluate our \guidedRetraining method and we assess its added value by comparing it to the base classifiers and the classic retraining method.
\begin{itemize}
    \item \textbf{RQ3:} How effective is \guidedRetraining in improving the classification on the difficult subset?
\end{itemize}

\subsection{Evaluation subjects}\label{sec:evaluation_setup:subjects}
To evaluate the effectiveness of our approach in boosting the performance of the base classifiers, we conduct our experiments on classifiers trained to detect Android malware.
Specifically, we apply our method on four state-of-the-art Android malware detectors from the literature. 
These detectors have been successfully replicated~\cite{reproduction} in a study that has considered malware detectors from leading venues in security, software engineering, and machine learning.
In the following, we present an overview of our four evaluation subjects:

\subsubsection{\drebin~\cite{drebin}}
In 2014, \drebin was presented at NDSS as a static analysis-based malware detector.
The effectiveness of \drebin has made it very popular in the field since it has been studied or experimentally compared to several works~\cite{annamalai, feng2016automated, deep_dive_drebin, chen2018tinydroid, Abaid}. 

This approach relies on various features that are extracted from the DEX and the manifest files.
Specifically, \drebin considers eight categories of features: used permissions, requested permissions, app components, filtered intents, hardware components, restricted API calls, network addresses, and suspicious API calls.
The extracted features are fed to a linear SVM classifier so it can learn to differentiate between malware and benign apps.

\subsubsection{\mama~\cite{mamadroid:ndss}}
This malware detector was presented in 2017 at NDSS. It aims to capture the behaviour of Android apps using Markov Chains.
\mama first generates the apps' call graphs and abstracts each API call either to the package name (i.e., this mode of abstraction is referred to as \mamaP variant) or the first component of the package name (i.e., \mamaF variant).
Then, Markov chains are leveraged to create the feature vectors by considering the abstracted API calls as the states and the probabilities for changing the states as the transitions.
The two variants of \mama train a Random Forest classifier (RF) with the generated feature vectors.

\subsubsection{\reveal~\cite{revealdroid}} 
In 2018, \reveal was published in the TOSEM journal as an obfuscation-Resilient malware detector.
This approach relies on static analysis to extract three types of features: Android-API usage, reflection-based features, and calls from native binaries.
Similarly to \drebin, the extracted features are embedded into vectors and used to train a linear SVM algorithm. 

\subsubsection{\malscan~\cite{malscan}}
This approach was presented in 2019 at the ASE conference.
It proposes to consider the call graph of an app as a social network and conduct centrality analyses on that graph.
\malscan contributes with four variants that are denoted after the name of the centrality measure used to generate the features set: \malscanD, \malscanH, \malscanK, and \malscanCL.
Two other variants are also adopted by \malscan, which are \malscanA and \malscanCO.
These two variants rely on feature sets that are the average or the concatenation of feature sets from the previous four variants.
\malscan's six classifiers are all trained with a 1-Nearest Neighbour algorithm.

\subsection{Dataset}\label{sec:evaluation_setup:dataset}
We conduct our experiments on a public dataset of Android malware and benign apps from the literature.
It has been collected from \az~\cite{androzoo}, which is a growing collection that contains more than 19 million apps crawled from different markets, including Google Play.
In this dataset, benign apps are defined as apps that have not been flagged by any antivirus engine from VirusTotal\footnote{\url{https://www.virustotal.com}}.
A sample is labelled as malware in the dataset if it is flagged by at least two antivirus engines.
The apps in this dataset are created between 2019 and 2020 (i.e., according to their compilation date).
In total, the dataset contains \num{78002} malware and \num{187797} benign apps.

\subsection{\texttt{Model} and auxiliary classifier architectures}\label{sec:evaluation_setup:architectures}
In this section we present the neural network architecture we adopt for the \texttt{Model} and the auxiliary classifier, which are both based on the multi-layer perception (MLP).

\subsubsection{The \texttt{Model}}
As stated in Section~\ref{sec:approach}, we use \texttt{Model} to refer to the Encoder and the Projection Network, that we train to generate contrasted embeddings of the difficult samples.
For the encoder, our MLP contains five fully connected layers that have 2048, 1024, 512, 256, and 128 neurons respectively.
The outputs from each layer are normalised and passed through a RELU activation function.
The size of the input in the Encoder is not fixed since it depends on the size of the feature vectors of each approach. 

For the Projection Network, we use a two layers MLP that receives normalised inputs from the Encoder.
The first layer has 64 neurons with a RELU activation function and the output layer contains 32 neurons.
After it is trained, only the embeddings at the last layer of the Encoder are considered~\cite{supContrastive}.

\subsubsection{The auxiliary classifier}
This neural network is used to classify the samples using the embeddings generated by the \texttt{Model}.
It contains five layers with 64, 32, 16, 8, and 2 neurons respectively.
The RELU activation function is applied to the normalised output of the first four layers.
Since we conduct our experiments on binary classifiers, the last layer contains two neurons with a Sigmoid activation function (i.e., to output prediction probabilities for the two classes).

\subsection{Experimental setup}\label{sec:evaluation_setup:setup}
We conduct our experiments using the PyTorch\footnote{\url{https://pytorch.org}} and scikit-learn\footnote{\url{https://scikit-learn.org}} libraries.
For the base classifiers training step (i.e., Section~\ref{sec:approach:step1}), we split the dataset into training (80\%), validation (10\%), and test (10\%), and we rely on the implementation of the evaluation subjects from the replication study~\cite{reproduction}.
We also set the percentage of FPs and FNs tolerated in the easy dataset (i.e., the values of the parameters $X$ and $Y$ described in Section~\ref{sec:approach:step2}) to $5\%$.
This value has been validated empirically to minimise the errors on the difficult datasets.

For training the \texttt{Models} and the auxiliary classifiers, we leverage a publicly available implementation\footnote{\url{https://github.com/HobbitLong/SupContrast}} of \contrastive. 
We set 2000 as a maximum number of epochs, and we stop the training if the optimised metric (i.e., the loss for the \texttt{Model} and the accuracy for the auxiliary classifier) does not improve after 100 epochs.
We also set the batch size to the number of training samples divided by 10.
Due to the huge size of the input vectors of some evaluated approaches, we had to divide their training size by 20, so the dataset could fit into the memory.
For the learning-rate hyper-parameter, we set its value to 0.001.

Since the evaluated subjects have different feature vector sizes and leverage different base classifier algorithms, we had to resolve some issues faced during our experiments which are related to:

\subsubsection{The size of the input vectors}\label{sec:evaluation_setup:setup:size_inputs}
We present in the first column of Table~\ref{tab:size_inpt_data} the size of the feature vectors in the difficult datasets of our evaluation subjects.
As we can see, \drebin and \reveal leverage huge input vectors that would need massive memory resources to conduct the training.
To solve this issue, we rely on feature selection methods to select the top best \num{200000} features for both \drebin and \reveal.
Though the performance might decrease when discarding the other features, this method can guarantee that the training is feasible.

\begin{table*}[ht]
	\begin{center}
		\caption{Size of input vectors, the number of samples and the number of FPs and FNs in the test subsets}
		\label{tab:size_inpt_data}
		\scalebox{1}{
			\begin{tabular}{c c |  c c | c c |  c c | c c |  c c }
			&  \multirow{3}{*}{\makecell{Size of input \\vectors in the  \\difficult datasets}} & \multicolumn{4}{c|} {\makecell{Number of samples in the test dataset\\(i.e., benign:\num{18739} and malware: \num{7841})}} &
			\multicolumn{6}{c} {Number of FPs and FNs in the test dataset}	\\ \cline{3-12}
				 
		     &   & \multicolumn{2}{c|} {Easy dataset} & \multicolumn{2}{c|} {Difficult dataset} & \multicolumn{2}{c|} {Whole dataset} & \multicolumn{2}{c|} {Easy dataset} & \multicolumn{2}{c} {Difficult dataset} \\
			 \cline{3-12}
			 
			 &   & benign & malware & benign & malware & FPs & FNs & FPs & FNs & FPs & FNs \\

			    \hline
				\drebin & \num{1184063} & \num{7824} & \num{4146} &  \num{10915} &  \num{3695} & \num{154} & \num{419}& \num{7}& \num{26}& \num{147}& \num{393} \\
			
				\reveal & \num{7882350} & \num{6120} & \num{5308} & \num{12619}  & \num{2533} & \num{90} & \num{625}& \num{2}& \num{37}& \num{88}& \num{588}\\
			
				\mamaF  &      \num{65} & \num{7770} & \num{743} &  \num{10969} & \num{7098} & \num{52} & \num{734}& \num{3}& \num{33}& \num{49}& \num{701}\\
				
				\mamaP   & \num{198916} & \num{9642} & \num{178} & \num{9097}  & \num{7663} & \num{136} & \num{322}& \num{4}& \num{10}& \num{132}& \num{312}\\
	
				\malscanD & \num{21986} & \num{10966} & \num{174} & \num{7773}  &  \num{7667}& \num{249} & \num{275}& \num{13}& \num{19}& \num{236}& \num{256}\\
			
				\malscanH & \num{21986} & \num{12863} & \num{206} & \num{5876}  & \num{7635} & \num{147} & \num{321}& \num{16}& \num{27}& \num{131}& \num{294}\\
				
				\malscanK & \num{21986} & \num{12025} & \num{190} &  \num{6714} &  \num{7651}& \num{212} & \num{430}& \num{12}& \num{23}& \num{200}& \num{407}\\
				
				\malscanCL & \num{21986} & \num{11133} & \num{174} & \num{7606}  & \num{7667} & \num{213} & \num{252}& \num{12}& \num{19}& \num{201}& \num{233}\\
				
				\malscanA & \num{21986} & \num{12100} & \num{186} & \num{6639}  &  \num{7655}& \num{243} & \num{420}& \num{16}& \num{16}& \num{227}& \num{404}\\
				
				\malscanCO & \num{87944} & \num{12045} & \num{190} & \num{6694}  &  \num{7651}& \num{187} & \num{414}& \num{13}& \num{16}& \num{174}& \num{398}\\
		\end{tabular}}
		\end{center}
\end{table*}

\subsubsection{The probabilities of prediction}\label{sec:evaluation_setup:setup:probas}
As we have mentioned in Section~\ref{sec:approach:step1}, our method requires a base classifier that can output probabilities of prediction.
This requirement is satisfied for \mama variants since the base classifier is Random Forest.

For \drebin and \reveal, they train a linear SVM algorithm that outputs a decision function (i.e., its absolute value indicates the distance of each sample to the hyper-plan that separates the two classes).
This decision function that we denote $f$ can take negative and positive values, and it is unbounded (i.e., it can take any value).
In our experiments, we apply a transformation on the decision function to obtain prediction probabilities:

\[  p_i =  \frac{f_i - f_\mathrm{min}}{f_\mathrm{max} - f_\mathrm{min}} \times (p_\mathrm{max} - p_\mathrm{min}) + p_\mathrm{min} \]

where $f_i$, $f_\mathrm{min}$, $f_\mathrm{max}$, $p_\mathrm{min}$, and $p_\mathrm{max}$ refer to the decision function value of sample $i$, the minimum and maximum values of $f$ and  the minimum and maximum values of $p$ respectively.
Basically, the transformation converts the positive values of the decision function into probabilities that are equal or greater than 0.5 and the negative values to probabilities smaller than 0.5.

As for \malscan variants, they rely on the 1-Nearest Neighbour classifier that outputs either 0 or 1 prediction probabilities.
Specifically, when the $k$ hyper-parameter of KNN is set to 1, each sample is labelled after its closest training sample in the dimensional space.
If the label of the nearest neighbour is positive, the sample receives the same label with a prediction probability of $p = 1$.
Otherwise, the sample is predicted as negative with $p = 0$.
While increasing the $k$ hyper-parameter can certainly widen the range of the prediction probabilities, we did not opt for this solution since it would change the configuration of the approach.

Since the aim of using the prediction probabilities is to identify the difficult samples, we resolve the issue of \malscan variants differently.
Specifically, we train a RF classifier to learn the correct predictions and the errors made by the 1-NN classifier.
If a sample has been misclassified by 1-NN, it is labelled as positive for RF training.
Otherwise, it takes a negative label.
This method has resulted in a high number of true negatives (i.e., a negative sample in the context of RF training means a sample that the KNN has correctly classified).
Since the number of negative samples is generally higher than the number of positives (i.e., errors made by KNN), RF has generated many false negative predictions.
Nevertheless, by selecting the samples that are predicted with a probability $p = 0$ by RF, we are able to identify a dataset that contains many true negatives and only few false negatives.
Translated to KNN predictions, this dataset contains many samples that are easy to predict and few difficult samples.
We consider this dataset as the easy subset, and all the other samples as the difficult subset in the case of \malscan variants.

\section{Evaluation results}\label{sec:evaluation_results}

\subsection{\textbf{RQ1:} To what extent it is feasible to split a dataset into two subsets, one with fewer prediction errors and one with most errors?}

In this section, we investigate the possibility to identify and separate the difficult samples in a dataset.
Specifically, we attempt to split up a given dataset into two subsets: easy and difficult subsets.
Most of the samples in the easy subset would be correctly classified by the base classifier (i.e., the easy dataset would contain only a few prediction errors).
As for the difficult subset, it would contain most of the prediction errors from the dataset we attempt to split.

We conduct our experiments on the evaluation subjects introduced in Section~\ref{sec:evaluation_setup:subjects}.
For \mama variants, we directly apply our method described in Section~\ref{sec:approach:step2} since the base classifiers output prediction probabilities.
For \drebin and \reveal, we use the technique described in Section~\ref{sec:evaluation_setup:setup:probas} to map the decision function values returned by the base classifiers (i.e., linear SVM) to prediction probabilities.
As for \malscan variants, the 1-NN base classifier does not output usable prediction probabilities (i.e., the probabilities are either 0 or 1).
We thus rely on the method described in Section~\ref{sec:evaluation_setup:setup:probas} for splitting the datasets.
We note that most classifiers described in scikit-learn documentation\footnote{\url{https://scikit-learn.org/stable/modules/classes.html}} generate prediction probabilities or decision function values. Consequently, when the base classifier does not directly output the prediction probabilities, our approach is still feasible using the techniques described in Section~\ref{sec:evaluation_setup:setup:probas}.

\begin{figure*}[!htbp]
    \centering
    \includegraphics[width=\linewidth, height=0.95\textheight]{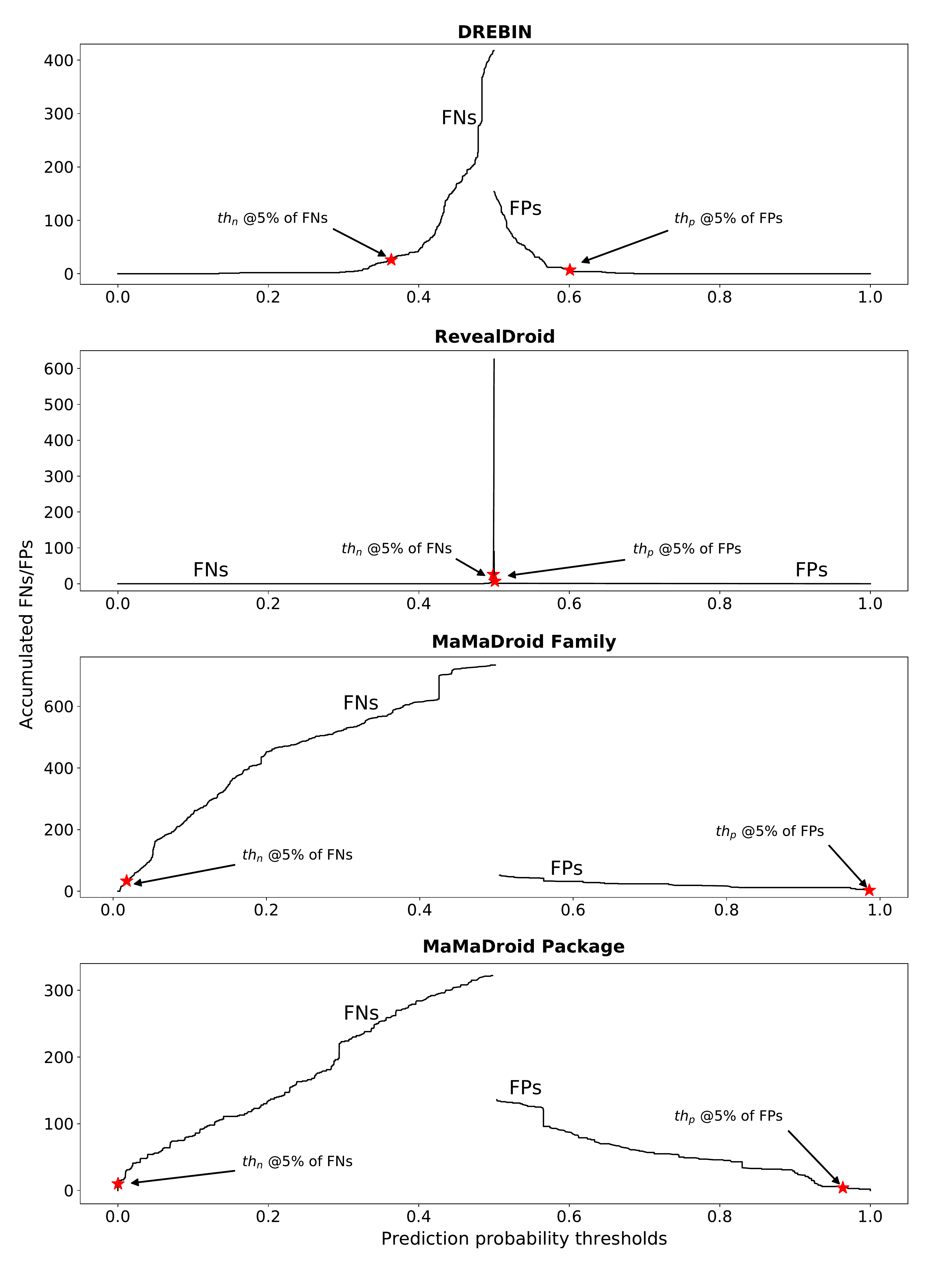}
    \caption{The accumulated number of FPs and FNs as a function of the prediction probability thresholds}
    \label{fig:errors_probas}
\end{figure*}

We report in Table~\ref{tab:size_inpt_data} the size of the easy and difficult subsets as well as the prediction errors made by the base classifier in each subset.
Overall, we are able to split the test dataset into easy and difficult subsets for all the evaluation subjects.
Indeed, the easy subsets contain few FPs and FNs made by the base classifiers.
As for the difficult subsets, they include most of the prediction errors from the whole test dataset.

We also present in Figure~\ref{fig:errors_probas} the evolution of the accumulated FPs and FNs against the prediction probability thresholds on the test dataset.
The graphs that are defined for prediction probabilities smaller than 0.5 represent the accumulated FNs.
Similarly, the accumulated FPs are represented by the graphs that are defined for prediction probabilities greater than 0.5.

From the Figure, we observe that the accumulated FNs are positively correlated with the prediction probabilities.
As for the accumulated FPs, they are negatively correlated with the prediction probabilities.
These two observations support our splitting method since we select the easy samples from the two ends of the graphs, where the FNs and FPs are low.


\highlight{
\textbf{RQ1 answer:} The difficult samples in a dataset can be identified and separated. Indeed, a dataset can be split into an easy subset with few prediction errors, and a difficult subset that contains most of the misclassifications made by the base classifier.}

\subsection{\textbf{RQ2:} How effective is the classic retraining method in improving the classification on the difficult subset?}

As we have seen in the previous section, we have created easy and difficult subsets based on the predictions of the base classifiers.
We can directly predict the class of the easy samples using the base classifiers since they make few classification mistakes on these samples.
For the difficult subsets, the prediction errors are important.

\begin{figure}[H]
    \centering
    \resizebox{0.3\linewidth}{!}{
    \begin{tikzpicture}
\tikzstyle{arrowStyle}=[-latex]
\tikzset{node/.style={rounded corners, align=center,scale=1}}

    \begin{scope}[cm={1,0,0,1,(1.55, 0)}]
    
    \draw[black, very thick] (0, 0.4) ellipse (1.2 and 0.6);
    \draw (0.02, 0.5) node[]{\scriptsize Difficult training};
    \draw (0.02, 0.28) node[]{\scriptsize dataset};
    
    \draw[->,>=latex,line width=1] (.05, -.2) -- (.05, -1.);
    \draw (.55, -.5) node[]{\scriptsize Train};
    
    \draw [green, line width=1] (-.44, -1.) rectangle (.56, -1.8);
    \draw (.06, -1.35) node[]{\scriptsize Model };
    
    \draw[->,>=latex,line width=1] (.05, -1.8) -- (.05, -2.6);
    \draw (.75, -2.) node[]{\scriptsize Generate};
    \draw (.75, -2.25) node[]{\scriptsize embedding};
    
    \draw (.06, -2.75) node[]{\scriptsize Embedding};
    
    \draw[->,>=latex,line width=1] (.05, -2.9) -- (.05, -3.7);
    \draw (.55, -3.2) node[]{\scriptsize Train};
    
    \draw [red, line width=1] (-.44, -3.7) rectangle (.56, -4.6); 
    \draw (.06, -4.) node[]{\scriptsize Auxiliary };
    \draw (.06, -4.3) node[]{\scriptsize classifier };
    
    \end{scope}
\end{tikzpicture}}
    \caption{An illustration of the classic retraining method}
    \label{fig:classic_retraining}
\end{figure}
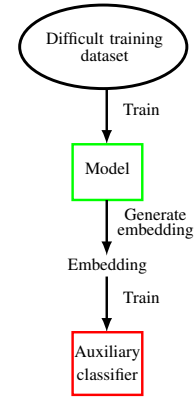

In this section, we investigate the impact of the classic retraining on the detection performance of the difficult samples.
In the classic retraining setting, we train only one \texttt{Model} to generate the embeddings of the difficult samples.
This training is conducted in one step using all the difficult samples in the training dataset.
Then, we directly train an auxiliary classifier on the generated embeddings. 
The classic retraining method consists of a trivial retraining that does not involve any guidance to generate the embeddings.
We present an illustration of the classic retraining method in Figure~\ref{fig:classic_retraining}

We define $\Delta Errors$ as the difference between the number of prediction errors made by the base classifier and the number of prediction errors from the auxiliary classifier on the difficult subsets. Its value can be positive or negative.
If it is positive, $\Delta Errors$ means that the auxiliary classifier has made more prediction errors than the base classifier.
If $\Delta Errors$ is negative, the auxiliary classifier has indeed improved the detection performance by decreasing the number of misclassifications reported by the base classifier.
Its formula is as follows:

\[  \Delta Errors = (FP_\mathrm{bc} + FN_\mathrm{bc}) - (FP_\mathrm{ac} + FN_\mathrm{ac}) \]

where $FP_\mathrm{bc}$, $FN_\mathrm{bc}$, $FP_\mathrm{ac}$, and $FN_\mathrm{ac}$ refer to the FPs and FNs of the base classifier (i.e., bc) and the auxiliary classifier (i.e., ac) respectively.

We calculate the accuracy, the F1-score, and the total number of prediction errors made by the base classifiers on the difficult samples, and we report their values in the first column (i.e., Base classifiers) of Table~\ref{tab:evaluation}.
We also report the accuracy and the F1-score of the classic retraining in the second column (i.e., Classic Retraining column) of Table~\ref{tab:evaluation}.
To quantitatively compare the prediction errors of the base classifiers and the classic retraining, we report the value of $\Delta Errors$ and the percentage of errors reduction for the different evaluation subjects in the second column of Table~\ref{tab:evaluation}.

\begin{table*}[ht]
	\begin{center}
		\caption{Comparison of the detection performance of the base classifiers, the auxiliary classifiers of \guidedRetraining, and the auxiliary classifiers of a simple retraining method on the difficult test dataset}
		\label{tab:evaluation}
		\resizebox{\linewidth}{!}{
			\begin{tabular}{c c c c |c c c  c |c c c c }
			     & \multicolumn{3}{c|} { Base classifiers} & \multicolumn{4}{c|} {\makecell{Classic Retraining \\ Auxiliary classifiers}} & \multicolumn{4}{c} {\makecell{\guidedRetraining \\ Auxiliary classifiers}} \\
				 \hline
				 &  A &  F1 & \makecell{\# Errors \\ (FPs + FNs)} & A &  F1 &  $\Delta Errors$ & \makecell{Errors \\ reduction} &  A &  F1 &  $\Delta Errors$ & \makecell{Errors \\ reduction} \\
			    \hline
				\drebin    & 96.3\% & 92.44\% & 540   & 96.74\% & 93.33\% & -63 & 11.67\%      &  \bf{96.78\%} & \bf{93.44\%} & -69 & 12.78\% \\
			
				\reveal   & 95.54\% & 85.19\% & 676   & 96.01\% & 87.18\% & -72 & 10.65\%       & \bf{97.18\%} & \bf{91.44\%} & -248 & 36.69\% \\
			
				\mamaF  & 95.85\% & 94.46\% & 750   & 97.13\% & 96.31\% & -232 & 30.93\%    & \bf{97.31\%} & \bf{96.55\%}  & -264 & 35.2\% \\
				
				\mamaP  & \bf{97.35\%} & \bf{97.07\%} & 444  & 97.2\% & 96.9\% & +25 & -5.63\%     & \bf{97.35\%} & \bf{97.07\%}  & 0 & 0\%\\
	
				\malscanD    & 96.81\% & 96.79\% & 492  & 97.47\% & 97.43\% & -104 & 21.14\%     & \bf{97.55\%} & \bf{97.51\%}  & -114 & 23.17\%\\
			
				\malscanH    & 96.85\% & 97.19\% & 425       & 96.84\% & 97.17\% & +2 & -0.47\%     & \bf{96.98\%} & \bf{97.29\%}  & -17 & 4\%\\
				
				\malscanK    & 95.77\% & 95.98\% & 607       & \bf{97.54\%} & \bf{97.67\%} & -254 & 41.85\%       & 97.47\% & 97.62\%  & -244 & 40.2\%\\
				
				\malscanCL    & 97.16\% & 97.16\% & 434     & 97.54\% & 97.54\% & -59 & 13.59\%    & \bf{97.56\%} & \bf{97.55\%}  & -61 & 14.06\%\\
				
				\malscanA    & 95.58\% & 95.83\% & 631  & 97.1\% & 97.26\% & -216 & 34.23\%      & \bf{97.37\%} & \bf{97.53\%}  & -255 & 40.41\% \\
				
				\malscanCO    & 96.01\% & 96.21\% & 572   & 94.91\% & 95.04\% & +158 & -27.62\%     & \bf{97.23\%} & \bf{97.39\%}  & -175 & 30.59\%\\
		\end{tabular}}
		\end{center}
\end{table*}

Overall, the classic retraining method improves the detection performance of seven out of ten base classifiers.
However, the value of $\Delta Errors$ shows that the classic retraining method has generated more prediction errors than the base classifiers in three cases.
For \malscanCO, the detection performance of the base classifier has remarkably decreased after the classic retraining that has increased the prediction errors by 27.62\%.

\highlight{
\textbf{RQ2 answer:} The classic retraining method does not improve the detection performance of all the base classifiers.
}

\subsection{\textbf{RQ3:} How effective is \guidedRetraining in improving the classification on the difficult subset?}
We have shown in the previous section that a simple retraining method (i.e., classic retraining) is not always sufficient to improve the detection performance on the difficult samples.
These samples seems to be challenging and need more advanced techniques for their classification.

In this section, we aim to assess the added value of \guidedRetraining on improving the classification of the difficult samples.
To this end, we train the \texorpdfstring{\texttt{Model}\textsubscript{5}}{} on the difficult subsets of our evaluation subjects to generate the embeddings as we have described in Section~\ref{sec:approach:step3}.
Then, we train the auxiliary classifiers using these embeddings to make the final classification.

We calculate the accuracy and the F1-score of \guidedRetraining auxiliary classifiers on the difficult samples, and we report their values in the third column (i.e., \guidedRetraining column) of Table~\ref{tab:evaluation}.
We also report the value of $\Delta Errors$ (i.e., it compares the classification errors of the base classifiers and \guidedRetraining auxiliary classifiers) and the errors reduction in the same Table.

We observe that \guidedRetraining has improved the detection performance of nine out of ten base classifiers. For \mamaP, the base and the auxiliary classifiers have both reported the same accuracy and F1-score values.
The gain in the detection performance can also be quantified using the $\Delta Errors$ metric.
In the worse case (i.e., the case of \mamaP classifier), our auxiliary classifier has generated the same number of misclassifications as the base classifier.
The other experiments show that \guidedRetraining corrects up to 264 prediction errors made by the base classifiers.
Furthermore, it reduces more than 30\% of the prediction errors for five base classifiers.

Compared to the classic retraining method, the accuracy and the F1-score of \guidedRetraining are higher in nine out of ten experiments.
For \malscanK, the classic retraining method has resulted in a slightly better F1-score.
The $\Delta Errors$ also demonstrates that \guidedRetraining can correctly classify more samples than the classic retraining method (i.e., up to 333 samples in the case of \malscanCO).

\highlight{
\textbf{RQ3 answer:} \guidedRetraining boosts the detection performance of the base classifiers on the difficult samples.
Indeed, it has reduced the prediction errors made by the base classifiers by up to 40.41\% .
Furthermore, \guidedRetraining generally results in higher detection performance than the classic retraining method.}

\section{Related Work} \label{section:related}
\subsection{The concept of difficult samples}
The notion of difficult or hard samples has been discussed in several previous works.
Researchers have attributed different definitions to this concept depending on its use case.
A study~\cite{tian-etal-2021-embedding} has defined the difficult samples in the context of data imbalance as the samples that belong to the minority class and overlap with the majority class in the embedding space.
The authors of~\cite{tian-etal-2021-embedding} have proposed a framework MISO that creates non-overlapping embeddings for the difficult samples based on anchor instances.
ADASYN~\cite{4633969} is an algorithm that helps learning from imbalanced datasets by focusing more on the difficult samples during synthetic data generation.
Specifically, ADASYN relies on a weighted distribution of the minority classes to generate the synthetic samples.
Adaboost~\cite{FREUND1997119} is an ensemble learning technique that combines the predictions of a series of base learners.
The basic idea of this technique is that each algorithm in the series increases the weights associated with the hard samples (i.e., samples that are incorrectly predicted) reported by the previous learner.

The Focal Loss~\cite{Lin_2017_ICCV} has been proposed to put more focus on the hard samples during the training.
The paper defines hard samples as instances on which the prediction error is high.
The Focal Loss modifies the Cross-Entropy loss by decreasing the loss weights of the samples that are correctly predicted.
Similarly, Dice Loss~\cite{li-etal-2020-dice} has also been proposed to equally deal with false positives and false negatives and mitigate the problems related to class imbalance.
This method assigns weights to each training sample. These weights dynamically decrease for the easy negative samples during the training.

In object detection field, Online Hard Example Mining (OHEM) algorithm~\cite{Shrivastava_2016_CVPR} has been proposed to automatically sample hard example during the training. This method modifies the Stochastic Gradient Descent by selecting diverse samples that have large losses in order to train region-based convolution networks.
Another approach~\cite{5255236} has also been proposed to represent highly variable classes using discriminative training~\cite{katagiri1999discriminative, JIANG2010589}. This paper defined hard instances as the samples that are incorrectly predicted by the classifier.

The notion of difficult samples has also been implicitly used in GANs~\cite{NIPS2014_5ca3e9b1}.
Specifically, the method relies on two models: a generator and a discriminator.
The generator is trained to produce adversary samples that are difficult to classify.
As for the discriminator, it is trained to distinguish between the samples originating from the dataset and the ones that are produced by the generator.

Our work differs from these related works by defining difficult samples as the instances that a base classifier is not very confident about their class labels, i.e.,  instances that do not receive high prediction probabilities by the base classifier.


\subsection{Retraining ML models}
Retraining is a technique that generally aims to improve the detection performance of the model. It has been defined and adopted in various ways in the literature.
DeltaGrad~\cite{pmlr-v119-wu20b} is proposed to retrain a model by updating its parameters after adding or deleting a set of training instances.
A Neural Network Tree algorithm~\cite{939114} has been proposed, which relies on a retraining technique that updates the weights of the neural networks.
The method iterates over all the training samples to minimise the prediction errors.
Similarly, retraining using predicted prior time series data has been proposed to improve the prediction of Anaerobic digestion~\cite{PARK2021117250}.
SURE~\cite{Feng_An_2019} is a partial label learning technique that is based on self-training.
It introduces the maximum infinity norm regularisation to generate pseudo-labels for the training samples.

In medical research, a method has been proposed to retrain the Epilepsy seizure detection model as more data becomes available~\cite{6091865}.
This retraining involves all the available data that is gradually collected based on feedback from patients or a
seizure detection module.
Similarly, daily new parameters measurements have been leveraged to retrain Gradient Boosting Tree algorithm to predict wind power~\cite{8471713}.
In IoT systems, a retraining technique~\cite{8888241} has been developed to train the model on both the original training dataset and the test set that has pseudo-generated labels.
Weighted Retraining~\cite{NEURIPS2020_81e3225c} is a method that updates the latent space with new instances and periodically retrains generative models (e.g., GANs~\cite{NIPS2014_5ca3e9b1}) to improve the optimisation.

Our \guidedRetraining method is intended to improve the classification on the difficult samples and is guided using the predictions of a base classifier.

\subsection{Android malware detection}
The literature of Android malware lavishes with diverse approaches that aim to detect malicious applications.
Many of the proposed detectors rely on ML and DL techniques as promising tools to achieve high detection performance.
In addition to the state of the art approaches that we have presented in Section~\ref{sec:evaluation_setup:subjects} of our experimental evaluation, we review in this section some other techniques that have contributed to Android malware detection.
PerDRaM~\cite{akbar2022permissions} is a malware detector that relies on permissions, smali size, and permissions rate features.
MLDroi~\cite{mahindru2021mldroid} detects malware using permissions, API calls, number of apps' download and apps' rating features.
TC-Droid~\cite{ZHANG2021107069} considers the apps analysis reports as text sequences and feeds them to a convolutional neural network model.
Besides, a multi-view malware detector~\cite{millar2021multi} has leveraged convolutional neural networks along with permissions, opcode sequences, and predefined Android API features.
Researchers have also contributed with studies that review the literature of Android malware detection~\cite{9130686, SHARMA2021100373, info12050185, liu2021deep} 

Recently, image-based Android malware detection has become popular due to its automatic features extraction~\cite{nadia2021dexray, Tiezhu2021, R2D2, ding2020android}.
This method involves no features engineering since the task of extracting the features is assigned to DL models.

With our \guidedRetraining method, we aim to enhance the detection performance of Android malware detectors and reduce their misclassifications.

\subsection{Supervised Contrastive Learning for malware detection}
Recently, a few studies for malware detection have leveraged \contrastive due to its promising results.
IFDroid~\cite{wu2021obfuscation} is an Android malware family classification approach that applies centrality analysis on the call graph of the apps and transforms them into images. \contrastive is then used to extract features from the images by considering the instances that belong to the same family as positive samples.
Malfustection~\cite{maghouli2021malfustection} is a malware classifier and Obfuscation detector that is based on semi-supervised contrastive learning.
The approach converts the program bytecode into an image and trains an encoder in an unsupervised way before fine-tuning it with labelled data.
CADE~\cite{263854} is a method to detect concept drift which has been evaluated on Android malware classification and network intrusion detection.
It leverages \contrastive to map the input samples into a low-dimensional space in which the distance between samples can be calculated for concept drift detection.
In our work, we leverage \contrastive to generate the embeddings of the difficult samples. This process is guided using the predictions of the base classifier. 

\section{Conclusion} \label{section:conclusion}
To evade detection, attackers devote time and effort to develop malicious software that resemble legitimate programs.
Consequently, many malware are difficult to distinguish from genuine programs, and thus manage to make their way into application markets.
Real-world software datasets are not perfectly separable into benign and malware samples due to the presence of malicious programs that are very similar to legitimate software and vice versa.
Indeed, these samples are challenging to malware detectors and require sophisticated techniques to achieve a high detection effectiveness. 

In this paper, we propose to split a binary dataset into subsets containing either easy or difficult samples.
The easy samples are efficiently predicted by a base classifier.
For the difficult samples, we propose a more advanced technique to better differentiate the two classes (malicious vs benign).
Specifically, we leverage Supervised Contrastive Learning to generate enhanced embeddings for the difficult input samples.
We rely on the predictions of the base classifier on the difficult samples to guide the retraining that generates the new representations.
Then, we train an auxiliary classifier on the new embeddings of the difficult samples.
We evaluate our method on four state of the art Android malware detectors using a large dataset of malicious and benign apps.
Our experiments show that \guidedRetraining boosts the detection performance on the difficult samples and reduces the prediction errors made by the base classifiers by up to 40.41\%.
We note that our method is not limited to Android malware detection and can be applied to other binary classification tasks.


\balance
\bibliographystyle{IEEEtranS}
\bibliography{biblio}

\end{document}